**Photon Echo in Uniaxially Stressed Germanium with Antimony Donors**

R.Kh. Zhukavin[a], V.D. Kukotenko[b], P.A. Bushuykin[a], Yu.Yu. Choporova[b], N.D. Osintseva[b], K.A. Kovalevsky[a], V.V. Tsyplenkov[a]*, V.V. Gerasimov[b,c], N. Dessmann[d], N.V. Abrosimov[e], V.N. Shastin[a]

[a] Institute for Physics of Microstructures, Russian Academy of Sciences, Nizhny Novgorod, 603950, Russia
[b] Budker Institute of Nuclear Physics, Siberian Branch of the Russian Academy of Sciences, Novosibirsk, 630090, Russia
[c] Novosibirsk National Research State University, Novosibirsk, 630090, Russia
[d] FELIX Laboratory, Radboud University, Nijmegen 6525 ED, The Netherlands
[e] Leibniz-Institut für Kristallzüchtung, 12489 Berlin, Germany
*Tsyplenkov1@yandex.ru

**Abstract.**

Photon echo is observed in *n*-type Ge uniaxially stressed along the [111] crystallographic direction, with a coherence relaxation time of 300 ps. The nonlinear polarization responsible for the effect originates from antimony donors. Uniaxial stress induces valley splitting of the donor states, substantially enhancing the coherence time and enabling the observation of photon echo.

**Introduction.**

Coulomb centers in semiconductors have long been the subject of intensive research due to a wide range of potential applications, from conventional micro- and nanoelectronic devices and terahertz sources (including lasers) to fundamentally new concepts such as single-atom transistors and quantum computing architectures. At present, technological advances have reached a level that enables the positioning and subsequent manipulation of individual impurity atoms in silicon and germanium crystals [1–6].

The feasibility and performance of devices based on Coulomb centers are, in many cases, determined by both the longitudinal ($T_1$) and transverse ($T_2$) relaxation times. In such systems, under conditions of low to moderate impurity concentrations ($N < 10^{16}$ cm$^{-3}$) and cryogenic temperatures ($T < 10$ K), the relaxation times $T_1$ and $T_2$ are governed by interactions with optical and acoustic phonons. Donor centers in germanium generally exhibit longer relaxation times than those in silicon [7,8].

An important feature of multivalley semiconductors is the possibility to control the point symmetry of wave functions and the energy spectrum of impurity states by applying uniaxial stress $S$. Modifying the symmetry and transition energies allows a substantial increase of both $T_1$ and $T_2$, which is not achievable in unstressed crystals. Long population and coherence lifetimes are essential for the efficient operation of various devices, including impurity-based lasers and amplifiers [9–11], as well as for the realization of key quantum-optical effects such as photon echo [12], Ramsey fringes [13], and electromagnetically induced transparency [14]. In addition, enhanced coherence improves the efficiency of nonlinear optical processes, including four-wave mixing and difference-frequency generation [15].

Among group-V donors in germanium, antimony exhibits the smallest ionization energy ($E_i \approx$ 10.45 meV) [16], which is close to the value predicted within the effective-mass approximation. The so-called chemical shift, determined by the valley–orbit interaction, is $\Delta_1 = 0.46$ meV ($T \approx$

5.8 K), which results in rapid thermal transitions $1s(A_1) \to 1s(T_2)$ and, consequently, fast coherence relaxation in the absence of uniaxial stress.

To efficiently modify the donor wave functions and energy spectrum, uniaxial stress applied along the [111] direction is the most favorable. Even at relatively small stress values exceeding $S_0 = \Delta_1/(\Xi_u c_{44}) \approx 50$bar (where $\Xi_u = 6.4$ eV is the deformation potential constant [17] and $c_{44} = 68.2$ GPa is the elastic constant of germanium [18]), the intervalley splitting $\Delta E$ exceeds the chemical shift $\Delta_1$. As a result, the donor states split into two groups: the lower one formed by contributions from a single conduction-band valley, and the other by contributions from the remaining three valleys [19].

These two groups of states interact only weakly with each other, since both radiative and nonradiative transitions are strongly suppressed due to the mismatch between the energy level separations and the energies of intervalley phonons. Consequently, for stress $S > S_0$, the donor ground state can be described as a single-valley state, while the first excited state corresponds to the $2p_0$ level (Fig. 1). This greatly simplifies the analysis of relaxation processes under resonant excitation and makes antimony donors in uniaxially stressed germanium a convenient and well-defined model system for studying interactions with coherent radiation.

The main objective of the present work is the observation of terahertz photon echo arising from the excitation of coherent superpositions of antimony donor states in uniaxially stressed germanium. The photon echo technique has previously been successfully applied to donor centers in silicon [12,20] for measuring coherence relaxation times $T_2$, demonstrating its high sensitivity and effectiveness.

**Experimental details and results**

The germanium crystal was grown by the Czochralski method with an antimony concentration of $10^{15}$ cm$^{-3}$ and a low compensation level of $10^{12}$ cm$^{-3}$. The sample dimensions were $2 \times 5 \times 7$mm$^3$. The $2 \times 5$mm$^2$ facet of the crystal was aligned with the [111] crystallographic direction. The angle between the two large polished faces was approximately 2°. Thermal contact between the sample and the cold finger was provided by indium foils. The sample was mounted in a custom-designed module that allowed the application of uniaxial stress. The module was installed in a continuous-flow helium cryostat (Janis ST-100) equipped with TPX windows transparent for wavelengths $\lambda > 15$ μm, and cooled to temperatures down to $T \approx 4$K.

To verify that the intervalley splitting $\Delta E$ exceeds the threshold value ($S > 100$ bar), at which the wave functions of the lowest donor states are essentially of single valley structure and the energy of the $1s(A_1) \to 2p_0$ transition becomes independent of further stress increase, the transmission spectra were measured. The linewidth of the $1s(A_1) \to 2p_0$ transition at half maximum, corresponding to the inhomogeneous broadening $\Delta\omega$, was approximately 0.2 meV. All temperatures reported below were measured by a sensor located at the base of the cold finger.

Excitation pulses were provided by a free-electron laser (FEL) at the Siberian Synchrotron and Terahertz Radiation Center [21], tunable in the wavelength range $\lambda = 90-340$ μm, with a pulse duration of $T_p \approx 100$ps and a repetition rate of 5.6 MHz. The spectral linewidth of the FEL was about 1% of the operating wavelength of 233 μm, corresponding to a donor transition energy of 5.32 meV. The beam diameter on the sample surface was approximately 2.5 mm. The electric-dipole matrix element of the $1s(A_1) \to 2p_0$ transition used to estimate the required field strength was taken from Ref. [22]. The average radiation power was 7 mW in the $\pi/2$-pulse channel and 28 mW in the $\pi$-pulse channel.

Photon echo (PE) measurements were performed using a standard technique (see, e.g., Ref. [12]). The polarizations of the $\pi/2$- and $\pi$-pulses were parallel to each other and to the stress axis. A detector (Golay cell) was positioned in the direction $\mathbf{k}_E = 2\mathbf{k}_2 - \mathbf{k}_1$, where $\mathbf{k}_1$ and $\mathbf{k}_2$ are the wave vectors of the $\pi/2$- and $\pi$-pulses, respectively, to measure the emitted intensity as a function of the delay time between the pulses. The $\pi$-pulse beam was incident normally to the sample surface, while the $\pi/2$-pulse was directed at an angle of 22° with respect to the $\pi$-pulse propagation direction. The longitudinal relaxation time $T_1$ was measured using a pump–probe technique [7,8] and was found to be 1.7 ns for the $1s(A_1) \rightarrow 2p_0$ transition under the same conditions.

A terahertz signal was detected in the $\mathbf{k}_E$ direction (Fig. 2). Blocking either the $\pi/2$ or $\pi$ excitation beam resulted in the disappearance of the signal. In addition, the detected intensity decreased monotonically with decreasing second-pulse power (Fig. 2a) and with increasing angular deviation $\alpha$ of the detector from the $\mathbf{k}_E$ direction (Fig. 2b). The dependence of the emitted signal on the delay time (Fig. 2c) exhibited the characteristic asymmetric shape of photon echo [23,24]: the signal was stretched for positive delays $\tau > 0$, where the $\pi$-pulse came to the the crystal after the $\pi/2$-pulse. The signal reached its maximum amplitude and duration at 4 K, while at 20 K it was strongly reduced and did not exceed the laser pulse duration.

At 4 K, the decay of the photon echo signal with increasing delay time $\tau$ occurred on a timescale of approximately $75 \pm 5$ ps (Fig. 2c), which does not exceed the duration of either the $\pi/2$- or $\pi$-pulses. Therefore, to correctly describe the photon echo effect and to extract the transverse relaxation time $T_2$, theoretical modeling was performed using the pulse durations, relaxation times, and inhomogeneous broadening values comparable to the experimental parameters.

**Theoretical modeling**

The modeling was performed assuming a small optical thickness of the sample, such that the excitation pulses are only weakly modified during propagation through the crystal and all impurity centers within the interaction volume experience the same driving field. This allows the system to be treated using a perturbative expansion in the optical field. In the zeroth-order approximation, the field is determined solely by the external excitation pulses. The polarization induced by this field acts as a source of an additional field, which constitutes a correction to the zeroth-order solution.

In the present work, we restrict the analysis to the first-order approximation and assume that the field correction is proportional to the polarization induced in the medium by the sequence of $\pi/2$- and $\pi$-pulses. The wavefronts of the excitation pulses are assumed to be plane. Under these conditions, the intensity of the coherently emitted radiation at a given frequency and propagation direction defined by the wave vector $\mathbf{k}$ is proportional to the convolution of the following functions:

$$I(t, \mathbf{k}, \tau) \sim |P(t, \mathbf{k}; \tau)|^2 * \sigma(\mathbf{k})$$

Here, $P(t, \mathbf{k}; \tau)$ is the spatial Fourier transform of the total medium polarization within the considered two-dimensional geometry (with the excitation pulses propagating at an angle to each other), and $\sigma(\mathbf{k})$ is the Fourier transform of the spatial profile defining the active region of the medium in which the polarization is excited. The latter is determined by the thickness of the crystal and the diameters of the laser beams. The function $P(t, \mathbf{k}; \tau)$ depends on time, while the delay $\tau$ between the excitation pulses enters as a parameter. Since the signal is detected by a slow detector, the total observed signal Sig as a function of the delay time $\tau$ is given by the time integral:

$$Sig(\mathbf{k},\tau) = \int I(t,\mathbf{k},\tau)dt$$

The polarization $P(t,\mathbf{k};\tau)$ was obtained from the Bloch equations for slowly varying amplitudes (see, e.g., Refs. [23–25]). The detuning between the center frequency of the excitation pulses and the impurity transition was assumed to be zero. Figure 3 (top) shows the dependence of the photon echo (PE) signal on the delay time $\tau$, while the bottom panel presents the temporal profile of the PE intensity for different values of $\tau$.

A key feature of photon echo is the appearance of an emission signal along the $\mathbf{k}_E$ direction, whose maximum occurs at approximately $2\tau$ relative to the maximum of the $\pi/2$-pulse. According to the modeling results, this behavior becomes well pronounced when the delay time $\tau$ exceeds the pulse duration. As seen in Fig. 3, the PE signal is not a single short pulse with a duration on the order of $1/\Delta\omega$, but exhibits a more complex structure consisting of several peaks. Such behavior is characteristic of the regime $\frac{1}{\Delta\omega} < T_p$ [26]. When $\Delta\omega$ was reduced by an order of magnitude, the modeled PE signal became single-peaked and followed the shape predicted by the simple photon echo theory.

Both the experimental and theoretical dependences exhibit oscillatory rather than smooth behavior. The Fourier spectrum of these oscillations is determined by the frequency of the impurity transition $1s(A_1) \to 1s(T_2)$ and by the generalized Rabi frequency, which is governed primarily by the radiation frequency deviation arising from the phase mismatch of the field oscillations in the $\pi/2$- and $\pi$-pulses.

**Discussion**

The experimentally measured emission signal $Sig(\mathbf{k}_E,\tau)$ as a function of the delay time $\tau$ is in good agreement with the results of theoretical modeling obtained for the same ratios of $T_p$, $T_1$, $T_2$, and $\Delta\omega$ (Fig. 3a). The observed decay of the signal follows the photon echo intensity dependence predicted by the simple echo model [23,24], $I_{\text{echo}} \propto \exp(-4\tau/T_2)$, where $T_2$ is determined solely by irreversible decoherence processes. By comparing the experimental data with the calculations, we extract a coherence relaxation time of $T_2 \approx 300$ ps.

Under conditions of sufficiently strong uniaxial stress, where states associated with different conduction-band valleys in germanium become fully decoupled, irreversible coherence loss can be governed by two channels: direct relaxation of the $2p_0$ state to the ground state $1s(A_1)$, and thermally activated transitions to higher-lying states ($2p_0 \to 3p_0$, $2p_\pm$, etc.). Since the longitudinal relaxation time measured by the pump–probe method significantly exceeds the transverse one ($T_1 \gg T_2$), coherence loss is dominated by phonon absorption on transitions from the upper excited state $2p_0$ to higher states.

In Ref. [27], the coherence relaxation time $T_2$ for arsenic donors in germanium was estimated by accounting for nonradiative transitions from the upper excited state of a shallow donor to nearby states, yielding values in the range of 100–500 ps for temperatures between 4 and 30 K. Because the spectra of the higher excited states of shallow donors are nearly identical, these results are also applicable to antimony donors. Comparison with these theoretical estimates indicates an effective temperature of approximately 12 K, which may result from limited heat removal from the sample under the relatively high average optical power used in the experiment.

**Conclusion**

Photon echo was observed for antimony donors in germanium at cryogenic temperatures under uniaxial stress upon resonant excitation of the $2p_0$ state by a sequence of two pulses propagating at an angle to each other. Comparison between the experimental results and theoretical modeling allowed us to determine the transverse coherence relaxation time of antimony donor states in uniaxially stressed germanium, governed by interactions with lattice vibrations, to be $T_2 = 300 \pm 20$ ps.

The authors thank the Novosibirsk Free-Electron Laser operators for their assistance. The work was done at the shared research center SSTRC on the basis of the Novosibirsk FEL at BINP SB RAS. Research was supported by the state assignment of IPM RAS (No. FFUF-2024-0019).

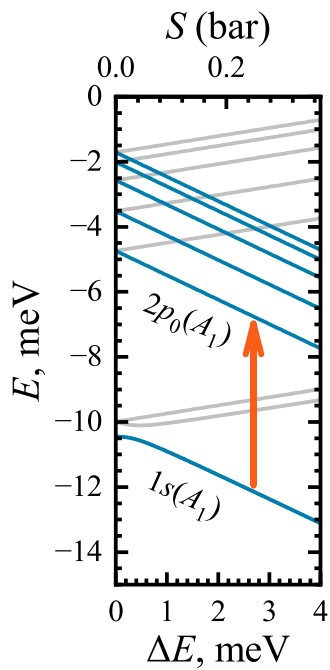

**FIG. 1.** Energy-level scheme of an Sb donor in Ge under uniaxial strain along the crystallographic [111] direction. The arrow indicates the transition used to create a coherent superposition of states.

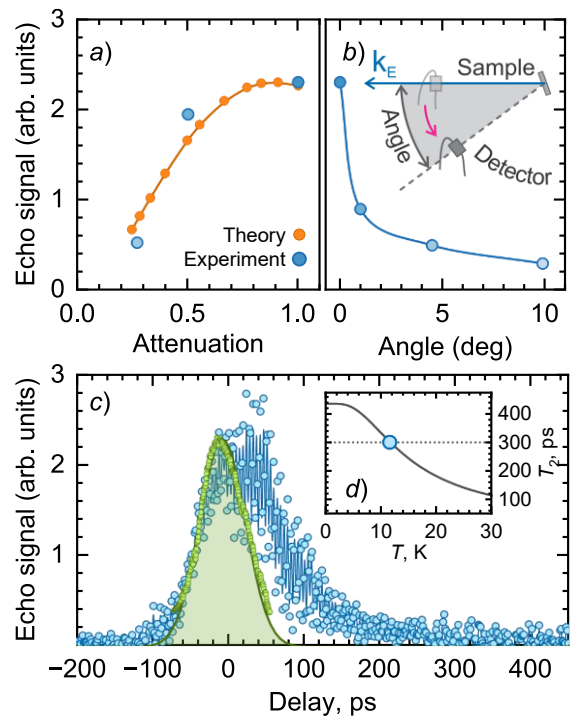

**FIG. 2.** Photon-echo signal amplitude in uniaxially strained Ge:Sb as a function of (a) the second-pulse intensity, (b) detector position, and (c) delay time between the $\pi/2$ and $\pi$ pulses, compared with the laser pulse. Inset (d) shows the extracted value $T_2 = 300$ ps placed on the theoretical temperature dependence of the coherence decay time from [27]. Experimental parameters: $P_\pi = 28$ mW, $P_{\pi/2} = 7$ mW, $\lambda = 232$ μm, $\mathbf{S} \parallel [111]$, $S > 100$ bar.

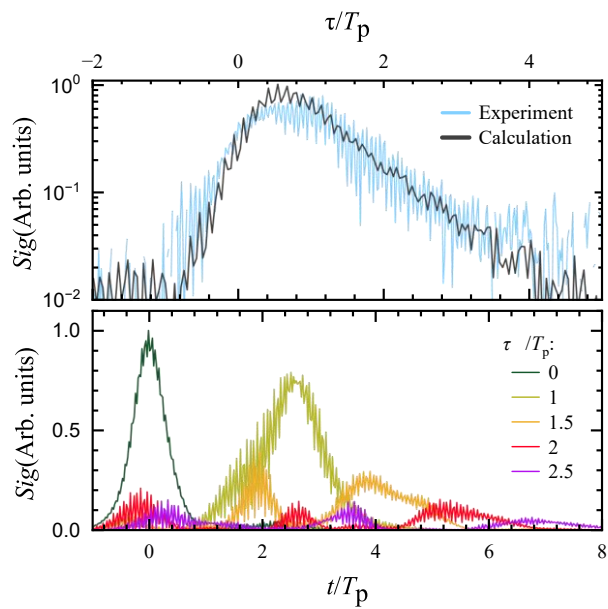

**FIG. 3.** (Top) Dependence of the photon-echo signal on the delay $\tau$, obtained experimentally (blue line) and simulated using the experimental parameters (gray line). (Bottom) Calculated time dependence of the PE intensity for different delays.